\documentclass[aps,pra,showpacs,twocolumn]{revtex4-1}
\usepackage{amssymb}
\usepackage{amsmath}
\usepackage{graphicx}
\usepackage{epsfig}
\usepackage{subfigure}
\usepackage{color}

\begin{document}

\title{Real-space decomposition of $p$-wave Kitaev chain}
\author{D. K. He, E. S. Ma}
\author{Z. Song} 
\email{songtc@nankai.edu.cn}
\affiliation{School of Physics, Nankai University, Tianjin 300071, China}

\begin{abstract}
We propose an extended Bogoliubov transformation in real space for spinless
fermions, based on which a class of Kitaev chains of length $2N$ with zero
chemical potential can be mapped to two independent Kitaev chains of length $%
N$. It provides an alternative way to investigate a complicated system from
the result of relatively simple systems. We demonstrate the implications of
this decomposition by a Su-Schrieffer-Heeger (SSH) Kitaev model, which
supports rich quantum phases. The features of the system, including the
groundstate topology and nonequilibrium dynamics, can be revealed directly
from that of sub-Kitaev chains. Based on this connection, two types of
Bardeen-Cooper-Schrieffer (BCS)-pair order parameters are introduced to
characterize the phase diagram, showing the ingredient of two different BCS
pairing modes. Analytical analysis and numerical simulations show that the
real-space decomposition for the ground state still holds true approximately
in presence of finite chemical potential in the gapful regions.
\end{abstract}

\maketitle

\affiliation{School of Physics, Nankai University, Tianjin 300071, China}

\section{Introduction}
The exact solution of a model Hamiltonian, especially for many-body system,
plays an important role in physics and sometimes may open the door to the
exploration of new frontiers in physics. One of common and efficient tools
is the Fourier transformation, which decomposes a Hamiltonian into many
commutative sub-Hamiltonians due to the translational symmetry. In contrast
to the $k$-space decomposition, there exist real-space decompositions, such
as the block-diagonalization based on the reflection symmetry and so on. In
this work, we propose another decomposition for the many-body Hamiltonian,
which bases on an extended Bogoliubov transformations in real space for
spinless fermions. By taking this transformation, we find that the
Hamiltonians of a class of Kitaev chains \cite{Kitaev} of length $2N$ with
zero chemical potential can be mapped to two commutative ones of Kitaev
chains of length $N$. All the eigen states can be constructed by the ones of
two sub-systems. It provides a clear physical picture and a simple way to
solve the dynamic problem in a complicated system from the result for
relative simple systems.\ We demonstrate the implications of this
decomposition by a SSH Kitaev model, which consists of dimerized hopping and
pairing terms and supports the rich quantum phases. The features of the
system, including the groundstate topology and nonequilibrium dynamics from
a trivial initial state, can be revealed directly from that of well-known
simple Kitaev chains. Based on this connection, two types of BCS-pair order
parameters are introduced to characterize the phase diagram, showing the
ingredient of two different BCS pairing modes.

For instance, the topological index of the original model can be shown to be
simply the sum of that of two sub-models. The ground state of an SSH Kitaev
model has rich quantum phases with winding numbers $\mathcal{N=}0$, $1$, and 
$2$, respectively, while a simple sub-system has quantum phases with winding
numbers $\mathcal{N=}0$ and $1$. The physical picture becomes clear in the
framework of decomposition.\ Based on the exact solution, the BCS-pair order
parameters, with respect to two independent sub-systems respectively, are
introduced to characterize the phase diagram by its value and nonanalytic
behavior at phase boundaries. We find that the ground state are two-fluid
condensate, with two different BCS pairing modes. In addition, such a
decomposition holds true not only for ground state but also the whole eigen
space. Then it should result in the decomposition of dynamics. Based on the
exact results on the dynamics obtained in the previous work\cite{SYB_PRB},
we study the dynamics of the SSH Kitaev model. The primary aim of this study
was to obtain pertinent information regarding the nonequilibrium state. In
this study, the starting point is the vacuum state, a trivial initial state.
We investigate the evolved states under the postquench Hamiltonian with
parameters covering the entire region and find two types of order
parameters that are different but can help determine the quantum phase
diagram. Finally, the robustness of such a decomposition to the perturbation
of nonzero chemical potential is also investigated. Analytical analysis and
numerical simulations show that the real-space decomposition for the ground
state still holds true approximately in many regions in the presence of
finite chemical potential.

This paper is organized as follows. In Sec. \ref{Model and decomposition},
we present a generalized model and introduce a real-space Bogoliubov
transformation to decompose the model Hamiltonian. In Secs. \ref{SSH Kitaev
model}, we apply the transformation on a SSH Kitaev model and deduce the
phase diagram from decomposed sub-Hamiltonians. In Secs. \ref{BCS order
parameters} and \ref{Non-equilibrium dynamics}, two types of BCS order
parameters are introduced. The explicit expressions for both ground state
and non-equilibrium state are obtained, respectively.\ Sec. \ref{Robustness
against nonzero} devotes to the investigation for the case with nonzero
potential.\ Finally, we provide a summary and discussion in Section \ref%
{sec_summary}.

\begin{figure*}[tbh]
\centering
\includegraphics[width=0.7\textwidth]{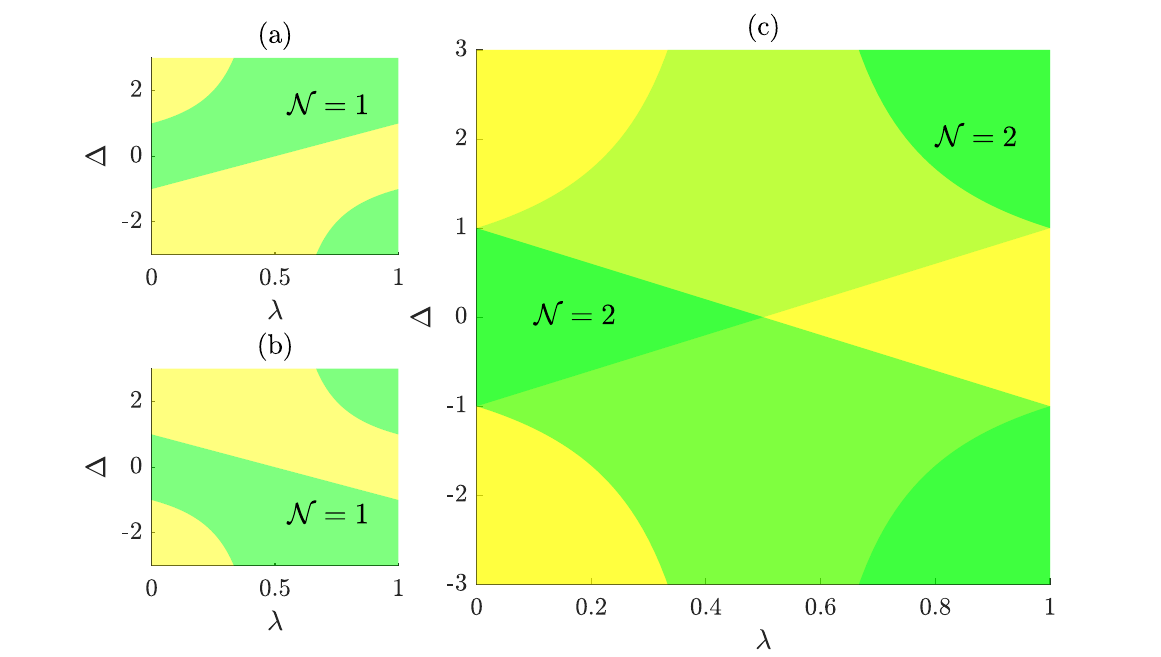}
\caption{Schematic of phase diagrams for the ground states of systems (a) $%
H_{A}$,\ (b) $H_{B}$ and (c) $H$, respectively. The phase boundaries
separating the regions with different colors are the plots of the curves
from Eqs. (\protect\ref{A1}), (\protect\ref{A2}), (\protect\ref{B1}), and (%
\protect\ref{B2}). The winding numbers are denoted in each regions. One can
see that all the information in (c) can be obtained directly by that of (a)
and (b), demonstrating the benefits of the real-space decomposition.}
\label{fig1}
\end{figure*}

\section{Model and decomposition}

\label{Model and decomposition}

We consider the following fermionic Hamiltonian on a lattice of length $2N$

\begin{equation}
H=\sum\limits_{j=1}^{2N}[J_{j}c_{j}^{\dag }c_{j+1}+\Delta _{j}c_{j}^{\dag
}c_{j+1}^{\dag }+\mathrm{H.c.}+\mu _{j}\left( 2n_{j}-1\right) ],  \label{H}
\end{equation}%
where $c_{j}^{\dag }$ $(c_{j})$\ is a fermionic creation (annihilation)
operator on site $j$, $n_{j}=c_{j}^{\dag }c_{j}$, $J_{j}$ the tunneling rate
across the dimer $\left( j,j+1\right) $, $%
%TCIMACRO{\U{3bc} }%
%BeginExpansion
\mu
%EndExpansion
_{j}$ the on-site chemical potential, and $\Delta _{j}$\ the strength of the 
$p$-wave pair creation (annihilation) on the dimer $\left( j,j+1\right) $. $%
c_{2N+1}=c_{1}$ is defined for periodic boundary condition. The Hamiltonian
in Eq. (\ref{H}) has a rich phase diagram that describes a spin-polarized $p$%
-wave superconductor in one dimension when the parameters $\left(
J_{j},\Delta _{j},%
%TCIMACRO{\U{3bc} }%
%BeginExpansion
\mu
%EndExpansion
_{j}\right) $\ are taken in a uniform fashion \cite{Ryohei Wakatsuki}. This
system has a topological phase in which a zero- energy Majorana mode is
located at each end of a long chain by taking the open boundary condition $%
c_{2N+1}=0$. It is the fermionized version of the well-known one-dimensional
transverse-field Ising model \cite{Pfeuty}, which is one of the simplest
solvable models when the translational symmetry is imposed that exhibits
quantum criticality and phase transition with spontaneous symmetry breaking 
\cite{SachdevBook,ZG}. In addition, several studies have been conducted with
a focus on long-range Kitaev chains \cite{DV1,DV2,OV,LL,UB}. Although system
we concerned in this work contains only the nearest neighbor hopping and
pairing terms, this method be applied to the system involving long-range
terms under certain conditions.

We introduce a linear unitary transformation

\begin{equation}
\left( 
\begin{array}{c}
\alpha _{j}^{\dag } \\ 
\alpha _{j} \\ 
\beta _{j}^{\dag } \\ 
\beta _{j}%
\end{array}%
\right) =T\left( 
\begin{array}{c}
c_{2j-1}^{\dag } \\ 
c_{2j-1} \\ 
c_{2j}^{\dag } \\ 
c_{2j}%
\end{array}%
\right) =T\left( 
\begin{array}{c}
A_{j}^{\dag } \\ 
A_{j} \\ 
B_{j}^{\dag } \\ 
B_{j}%
\end{array}%
\right) ,  \label{Tran}
\end{equation}%
with the matrix%
\begin{equation}
T=\left( T^{\dag }\right) ^{-1}=\frac{1}{2}\left( 
\begin{array}{cccc}
1 & 1 & 1 & -1 \\ 
1 & 1 & -1 & 1 \\ 
-i & i & i & i \\ 
-i & i & -i & -i%
\end{array}%
\right) ,
\end{equation}%
which ensures that the $\alpha _{j}$\ and $\beta _{j}$\ are still fermionic
operators, i.e., 
\begin{eqnarray}
\left\{ \alpha _{j},\alpha _{j^{\prime }}^{\dag }\right\} &=&\left\{ \alpha
_{j},\beta _{j^{\prime }}^{\dag }\right\} =\delta _{j,j^{\prime }},\left\{
\alpha _{j},\beta _{j^{\prime }}^{\dag }\right\} =0,  \notag \\
\left\{ \alpha _{j},\alpha _{j^{\prime }}\right\} &=&\left\{ \beta
_{j},\beta _{j^{\prime }}\right\} =\left\{ \alpha _{j},\beta _{j^{\prime
}}\right\} =0,
\end{eqnarray}%
and any transformed Hamiltonian maintains its original physics. It is
similar to the Bogoliubov transformation if regarding the parity of the site
number ($2j$ or $2j-1$) as spin degree of freedom. Applying the
transformation (\ref{Tran}), the Hamiltonian $H$ can be expressed in the form%
\begin{equation}
H=H_{A}+H_{B}+H_{AB},
\end{equation}%
where%
\begin{eqnarray}
&&H_{A}=\frac{1}{2}\sum\limits_{j=1}^{N}[\left( J_{2j}+\Delta _{2j}\right)
\left( \alpha _{j}^{\dag }\alpha _{j+1}+\alpha _{j}^{\dag }\alpha
_{j+1}^{\dag }\right)  \notag \\
&&+\mathrm{H.c.}+\left( \Delta _{2j-1}-J_{2j-1}\right) \left( 2\alpha
_{j}^{\dag }\alpha _{j}-1\right) ],
\end{eqnarray}%
and%
\begin{eqnarray}
&&H_{B}=\frac{1}{2}\sum\limits_{j=1}^{N}[\left( \Delta _{2j}-J_{2j}\right)
\left( \beta _{j}^{\dag }\beta _{j+1}+\beta _{j}^{\dag }\beta _{j+1}^{\dag
}\right)  \notag \\
&&+\mathrm{H.c.}+\left( J_{2j-1}+\Delta _{2j-1}\right) \left( 2\beta
_{j}^{\dag }\beta _{j}-1\right) ],
\end{eqnarray}%
denote two Kitaev chain of length $N$ and commute with each other%
\begin{equation}
\left[ H_{A},H_{B}\right] =0.
\end{equation}%
The third term is $\left\{ \mu _{l}\right\} $\ dependent, containing the
intra- and inter-chain interactions%
\begin{eqnarray}
H_{AB} &=&i\sum_{j=1}^{N}[\left( \mu _{2j-1}+\mu _{2j}\right) \beta
_{j}^{\dag }\alpha _{j}^{\dag }  \notag \\
&&+\left( \mu _{2j-1}-\mu _{2j}\right) \beta _{j}^{\dag }\alpha _{j}]+%
\mathrm{H.c.}
\end{eqnarray}%
Obviously, when consider the case with zero chemical potential, $\mu _{l}=0$%
, the Hamiltonian is exactly decomposed into two independent ones.
Accordingly we have 
\begin{equation}
H\left\vert \psi \right\rangle =H\left\vert \psi _{A}\right\rangle
\left\vert \psi _{B}\right\rangle =\left( \varepsilon _{A}+\varepsilon
_{B}\right) \left\vert \psi \right\rangle ,
\end{equation}%
with 
\begin{equation}
H_{A}\left\vert \psi _{A}\right\rangle =\varepsilon _{A}\left\vert \psi
_{A}\right\rangle ,H_{B}\left\vert \psi _{B}\right\rangle =\varepsilon
_{B}\left\vert \psi _{B}\right\rangle .
\end{equation}%
For arbitrary operator functions $F_{A}(\alpha _{1},...,$ $\alpha
_{l},...,\alpha _{N};$ $\alpha _{1}^{\dag },...,$ $\alpha _{j}^{\dag
},...,\alpha _{N}^{\dag })$ and $F_{B}(\beta _{1},...,$ $\beta _{l},...,$ $%
\beta _{N};$ $\beta _{1}^{\dag },...,$ $\beta _{j}^{\dag },...,\beta
_{N}^{\dag })$, we always have%
\begin{eqnarray}
\left\langle \psi \right\vert F_{A}\left\vert \psi \right\rangle
&=&\left\langle \psi _{A}\right\vert F_{A}\left\vert \psi _{A}\right\rangle ,
\notag \\
\left\langle \psi \right\vert F_{B}\left\vert \psi \right\rangle
&=&\left\langle \psi _{B}\right\vert F_{B}\left\vert \psi _{B}\right\rangle .
\end{eqnarray}%
Then any features relate to $F_{A}$($F_{B}$) in chain $A$($B$) should emerge
in the original system. The the following we will demonstrate this in a
concrete example.

\section{SSH Kitaev model}

\label{SSH Kitaev model} Now we consider a SSH Kitaev model with the
Hamiltonian

\begin{eqnarray}
&&H=\sum_{j=1}^{N}[\lambda c_{2j}^{\dag }c_{2j+1}+\left( 1-\lambda \right)
c_{2j-1}^{\dag }c_{2j}+\lambda \Delta c_{2j-1}c_{2j}  \notag \\
&&+\left( 1-\lambda \right) \Delta c_{2j}c_{2j+1}+2\mu \left(
n_{2j-1}+n_{2j}-1\right) ]+\mathrm{H.c.},
\end{eqnarray}%
in which, both the hopping and pairing strengths are assigned alternatively.
It has been well studied in the refs. \cite{Ryohei Wakatsuki}. In this work,
we focus on the connection between this $2N$-site system and two $N$-site
sub-systems.\ When taking $\mu =0$, the original Hamiltonian can be written
in the form $H=H_{A}+H_{B}$\ with

\begin{eqnarray}
H_{\sigma } &=&\sum_{j=1}^{N}[J_{\sigma }\left( d_{j,\sigma }^{\dag
}d_{j+1,\sigma }^{\dag }+d_{j,\sigma }^{\dag }d_{j+1,\sigma }\right) +\text{%
H.c.}  \notag \\
&&+\mu _{\sigma }\left( 1-2d_{j,\sigma }^{\dag }d_{j,\sigma }\right) ,
\end{eqnarray}%
where $\sigma =A$ or $B$, $d_{j,A}=\alpha _{j}$, $d_{j,B}=\beta _{j}$, and
the corresponding parameters are

\begin{eqnarray}
J_{A} &=&\left[ \lambda -\left( 1-\lambda \right) \Delta \right] /2, \\
\mu _{A} &=&\left[ \lambda \Delta +\left( 1-\lambda \right) \right] /2, \\
J_{B} &=&-\left[ \lambda +\left( 1-\lambda \right) \Delta \right] /2, \\
\mu _{B} &=&\left[ \lambda \Delta -\left( 1-\lambda \right) \right] /2.
\end{eqnarray}%
We note that both $H_{A}$ and $H_{B}$\ describe the same system but with
different parameters. Although the chemical potential is zero in $H$, it is
nonzero for sub-Hamiltonians. The phase diagram for each sub-Hamiltonian $%
H_{\sigma }$\ is well known, and then can be used to obtain the phase
diagram of $H$.\ In fact, the phase boundary of the ground state of $%
H_{\sigma }$\ are two points $\mu _{\sigma }/J_{\sigma }=\pm 1$. This maps
to two curves 
\begin{equation}
\Delta -2\lambda +1=0,  \label{A1}
\end{equation}%
and%
\begin{equation}
2\lambda \Delta -\Delta +1=0,  \label{A2}
\end{equation}%
in the $\lambda \Delta $-plane for $H_{A}$, while two curves%
\begin{equation}
\Delta +2\lambda -1=0,  \label{B1}
\end{equation}%
and%
\begin{equation}
2\lambda \Delta -\Delta -1=0,  \label{B2}
\end{equation}%
for $H_{B}$. According to our above analysis, all the four curves constitute
the phase diagram of $H$. In Fig. \ref{fig1}, we schematically illustrate
the phase diagrams of $H_{A}$,\ $H_{B}$ and $H$, respectively. We can see
that all the information in the ground state of $H$, including the gapless
lines and winding numbers can be obtained directly by that of $H_{A}$ and $%
H_{B}$, as a demonstration of benefits from the real-space decomposition.

\section{BCS order parameters}

\label{BCS order parameters}

It has been shown that the pairing order parameter can be utilized to
characterize the phase diagram \cite{SYB_PRB} and the long-range order \cite
{MES_PRB} of the ground state of a simple Kitaev chain. In this section, we
focus on the similar investigation in this aspect for the present model
based on the real-space decomposition method. We first briefly review the
obtained conclusion for the simple Kitaev chain. We take $H_{A}$\ as an
example. We introduce the BCS-pairing operator 
\begin{equation}
\widehat{O}_{A,k}=i\left( \alpha _{-k}\alpha _{k}-\alpha _{k}^{\dag }\alpha
_{-k}^{\dag }\right) ,
\end{equation}%
to characterize pairing channels in ${k}$ space. Here the Fourier
transformation%
\begin{eqnarray}
\alpha _{k} &=&\frac{1}{\sqrt{N}}\sum\limits_{j}e^{-ikj}\alpha _{j}  \notag
\\
&=&\frac{1}{2}\left( A_{k}+B_{k}+A_{-k}^{\dag }-B_{-k}^{\dag }\right) ,
\end{eqnarray}%
is applied with%
\begin{equation}
\left( 
\begin{array}{c}
A_{k} \\ 
B_{k}%
\end{array}%
\right) =\frac{1}{\sqrt{N}}\sum_{k}e^{-ikj}\left( 
\begin{array}{c}
A_{j} \\ 
B_{j}%
\end{array}%
\right) .  \label{Fourier}
\end{equation}%
For a given state $\left\vert \psi \right\rangle $, the quantity $%
|\left\langle \psi \right\vert \widehat{O}_{A,k}\left\vert \psi
\right\rangle |$\ measures the rate of transition for a BCS pair at the $k$\
channel. For the ground state $\left\vert \text{\textrm{G}}_{A}\right\rangle 
${\ of }$H_{A}${\ }the pairing order parameter of the ground state is
expressed as%
\begin{equation}
O_{A,\mathrm{g}}=\frac{1}{N}\sum_{\pi >k>0}\left\vert \left\langle \text{%
\textrm{G}}_{A}\right\vert \widehat{O}_{A,k}\left\vert \text{\textrm{G}}%
_{A}\right\rangle \right\vert .
\end{equation}%
In the large $N$ limit, it can be expressed explicitly as%
\begin{eqnarray}
O_{A,\mathrm{g}} &=&\frac{1}{\pi }\int_{0}^{\pi }\frac{\sin k}{2\sqrt{\left(
\mu _{A}/J_{A}-\cos k\right) ^{2}+\sin ^{2}k}}\mathrm{d}k  \notag \\
&=&\frac{1}{\pi }\left\{ 
\begin{array}{cc}
1, & \left\vert \mu _{A}/J_{A}\right\vert \leqslant 1 \\ 
J_{A}/\mu _{A}, & \left\vert \mu _{A}/J_{A}\right\vert >1%
\end{array}%
\right. .
\end{eqnarray}%
Obviously, there exist non-analytic points at $\left\vert \mu
_{A}/J_{A}\right\vert =1$, as the signatures of quantum phase boundary. The
same results hold for $H_{B}$.

\begin{figure*}[tbh]
\centering
\includegraphics[width=0.8\textwidth]{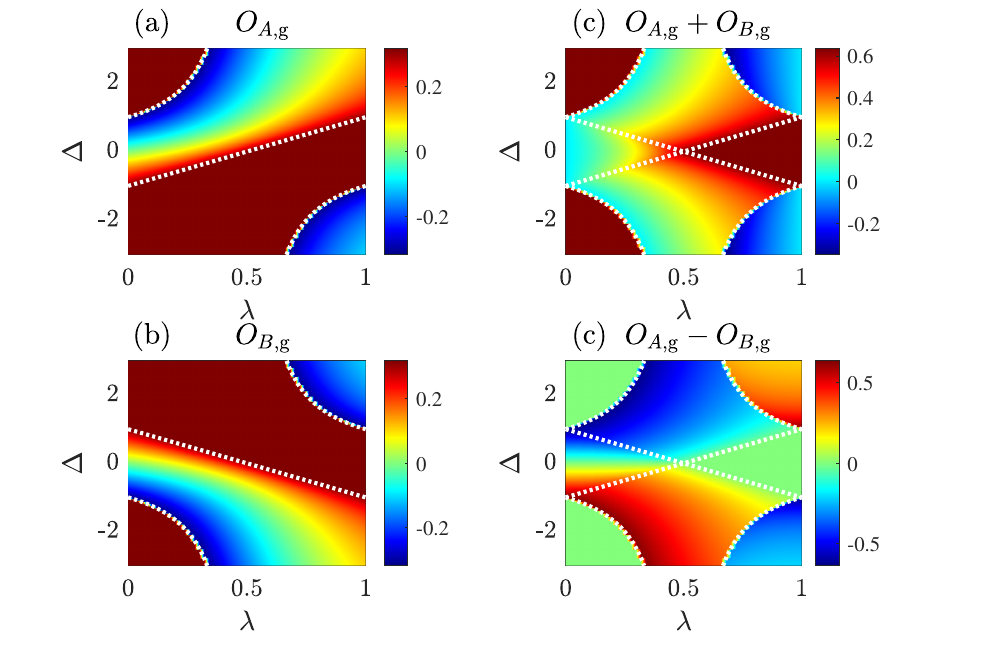}
\caption{Color contour plots of numerical results of order parameters (a) $%
O_{A,\mathrm{g}}$ and (b) $O_{B,\mathrm{g}}$\ defined in (\protect\ref{Oag})
and (\protect\ref{Obg}), respectively. (c) and (d) are plots of $O_{A,%
\mathrm{g}}+O_{B,\mathrm{g}}$ and $O_{A,\mathrm{g}}-O_{B,\mathrm{g}}$,
respectively. The system parameters are $N=1000$ and $J=1$. The white dashed
lines are a guide to the eye to indicate the phase boundaries presented in
Fig. \protect\ref{fig1}.\ It is clear that the order parameters obtained
from subsystems can identify the entire phase diagram.}
\label{fig2}
\end{figure*}

These conclusions can be directly applied on the ground state $\left\vert 
\text{\textrm{G}}\right\rangle $\ of system $H$, mapping the non-analytic
points $\left\vert \mu _{A}/J_{A}\right\vert =\left\vert \mu
_{B}/J_{B}\right\vert =1$ on the $\lambda \Delta $-plane. In fact,
introducing two order parameters%
\begin{equation}
O_{A,\mathrm{g}}=\frac{1}{N}\sum_{\pi >k>0}\left\vert \left\langle \text{%
\textrm{G}}\right\vert i\left( \alpha _{-k}\alpha _{k}-\alpha _{k}^{\dag
}\alpha _{-k}^{\dag }\right) \left\vert \text{\textrm{G}}\right\rangle
\right\vert ,
\end{equation}%
and%
\begin{equation}
O_{B,\mathrm{g}}=\frac{1}{N}\sum_{\pi >k>0}\left\vert \left\langle \text{%
\textrm{G}}\right\vert i\left( \beta _{-k}\beta _{k}-\beta _{k}^{\dag }\beta
_{-k}^{\dag }\right) \left\vert \text{\textrm{G}}\right\rangle \right\vert ,
\end{equation}%
we directly have the explicit expressions of $O_{A,\mathrm{g}}$\ and $O_{B,%
\mathrm{g}}$\ in the $\lambda \Delta $-plane%
\begin{equation}
O_{A,\mathrm{g}}=\frac{1}{\pi }\left\{ 
\begin{array}{cc}
1, & \text{regions I, II, and III} \\ 
\frac{\lambda -\left( 1-\lambda \right) \Delta }{\lambda \Delta +\left(
1-\lambda \right) }, & \text{otherwise}%
\end{array}%
\right. ,  \label{Oag}
\end{equation}%
where region I is $\Delta \in (-\frac{1}{2\lambda -1},\infty )$ for $%
0<\lambda <0.5$, region II is $\Delta \in (-\frac{1}{2\lambda -1},2\lambda
-1)$ for $0.5<\lambda <1$, and region III is $\Delta \in (-\infty ,2\lambda
-1)$ for $0<\lambda <0.5$, while%
\begin{equation}
O_{B,\mathrm{g}}=\frac{1}{\pi }\left\{ 
\begin{array}{cc}
1, & \text{regions I, II, and III} \\ 
-\frac{\lambda +\left( 1-\lambda \right) \Delta }{\lambda \Delta -\left(
1-\lambda \right) }, & \text{otherwise}%
\end{array}%
\right. ,  \label{Obg}
\end{equation}%
where region I is $\Delta \in (-\infty ,\frac{1}{2\lambda -1})$ for $%
0<\lambda <0.5$, region II is $\Delta \in (-2\lambda +1,\frac{1}{2\lambda -1}%
)$ for $0.5<\lambda <1$, and region III is $\Delta \in (-2\lambda +1,\infty
) $ for $0<\lambda <0.5$. In the representation of the original Hamiltonian,
the physics of the two order parameters are clear, based on the expressions%
\begin{eqnarray}
O_{A,\mathrm{g}}+O_{B,\mathrm{g}} &=&\frac{1}{N}\sum_{\pi >k>0}\left\vert
\left\langle \text{\textrm{G}}\right\vert \Lambda _{k}\left\vert \text{%
\textrm{G}}\right\rangle \right\vert , \\
O_{A,\mathrm{g}}-O_{B,\mathrm{g}} &=&\frac{1}{N}\sum_{\pi >k>0}\left\vert
\left\langle \text{\textrm{G}}\right\vert \Sigma _{k}\left\vert \text{%
\textrm{G}}\right\rangle \right\vert ,
\end{eqnarray}%
with the pairing and current operators%
\begin{eqnarray}
\Lambda _{k} &=&i\left( A_{-k}B_{k}+B_{-k}A_{k}-\text{\textrm{H.c.}}\right) ,
\\
\Sigma _{k} &=&i\left( A_{k}^{\dag }B_{k}+B_{-k}A_{-k}^{\dag }-\text{\textrm{%
H.c.}}\right) .
\end{eqnarray}%
Obviously, $O_{A,\mathrm{g}}+O_{B,\mathrm{g}}$\ measures the BCS pair for
fermions from two sub-lattices $A$ and $B$, while $O_{A,\mathrm{g}}-O_{B,%
\mathrm{g}}$\ the current across two sub-lattices. In Fig. \ref{fig2}, we
plot the functions, $O_{A,\mathrm{g}}$, $O_{B,\mathrm{g}}$\ and $O_{A,%
\mathrm{g}}\pm O_{B,\mathrm{g}}$, respectively. We can see that quantities $%
O_{A,\mathrm{g}}$ and $O_{B,\mathrm{g}}$\ reflect partial phase boundary,
while both two types of order parameters $O_{A,\mathrm{g}}\pm O_{B,\mathrm{g}%
}$\ can identify the entire phase diagram.

\section{Non-equilibrium dynamics}

\label{Non-equilibrium dynamics}

Recently, advancements in atomic physics, quantum optics, and nanoscience
have allowed the development of artificial systems with high accuracy \cite
{Jochim, Greiner}. The study of nonequilibrium many-body dynamics presents
an alternative approach for accessing a new exotic quantum state with an
energy level considerably different from that of the ground state \cite
{Choi,Else,Khemani,Lindner,Kaneko,Tindall,YXMPRA,ZXZPRB2,TK,JT1,JT2}.
Several works show that the quenching dynamics governed by the post-quench
Hamiltonian is intimately related to its ground state \cite{SYB_PRB,MH,LZ}.
In this section, we turn to the topic of nonequilibrium phenomena in the
present Hamiltonian. For a uniform Kitaev chain, it has been shown that the
pairing order parameter for a nonequilibrium state obtained through time
evolution from an initially prepared vacuum state still help determine the
phase diagram \cite{SYB_PRB}. For the present model, we will investigate
this issue based on the obtained results for the sub-systems $H_{A}$ and\ $%
H_{B}$.

Similarly, we still give a brief review the obtained conclusion for a simple
Kitaev chain, such as chain $H_{A}$. It has been shown that the order
parameter for a nonequilibrium state is defined as \cite{SYB_PRB}%
\begin{equation}
\overline{O}_{A}=\lim_{T\rightarrow \infty }\frac{1}{T}\int_{0}^{T}\frac{1}{N%
}\sum_{k}\left\vert \left\langle \psi (t)\right\vert \widehat{O}%
_{A,k}\left\vert \psi (t)\right\rangle \right\vert \mathrm{d}t,
\end{equation}%
where the time evolution $\left\vert \psi _{A}(t)\right\rangle
=e^{-iH_{A}t}\left\vert \psi _{A}(0)\right\rangle $ for a particular initial
state $\left\vert \psi (0)\right\rangle $, satisfying

\begin{equation}
\alpha _{j}\left\vert \psi _{A}(0)\right\rangle =0.
\end{equation}%
State $\left\vert \psi (0)\right\rangle $ is essentially an empty state for
a set of fermions $\left\{ \alpha _{j}\right\} $ in real space. In the large 
$N$ limit, it can be expressed explicitly as%
\begin{eqnarray}
\overline{O}_{A} &=&\frac{1}{2\pi }\int_{0}^{\pi }\left\vert \frac{\left(
\cos k-\mu _{A}/J_{A}\right) \sin k}{\left( \cos k-\mu _{A}/J_{A}\right)
^{2}+\sin ^{2}k}\right\vert \mathrm{d}k  \notag \\
&=&\frac{1}{2\pi }\left\{ 
\begin{array}{cc}
1, & \left\vert \mu _{A}/J_{A}\right\vert \leqslant 1 \\ 
\Lambda , & \left\vert \mu _{A}/J_{A}\right\vert >1%
\end{array}%
\right. ,
\end{eqnarray}%
where 
\begin{equation}
\Lambda =\left\vert \frac{J_{A}}{\mu _{A}}+\frac{1}{2}\left( 1-\frac{%
J_{A}^{2}}{\mu _{A}^{2}}\right) \ln \left\vert \frac{\mu _{A}+J_{A}}{\mu
_{A}-J_{A}}\right\vert \right\vert .
\end{equation}%
The non-analytic points at $\left\vert \mu _{A}/J_{A}\right\vert =1$ are the
evidently signatures of quantum phase boundary. The same results hold for $%
H_{B}$.

These conclusions can be directly applied to the dynamics\ of system $H$,
mapping the non-analytic points $\left\vert \mu _{A}/J_{A}\right\vert
=\left\vert \mu _{B}/J_{B}\right\vert =1$ on the $\lambda \Delta $-plane.
Accordingly, the nonequilibrium order parameters are defined for the time
evolution $\left\vert \psi (t)\right\rangle =e^{-iHt}\left\vert \psi
(0)\right\rangle $ $=e^{-iH_{A}t}e^{-iH_{B}t}\left\vert \psi
(0)\right\rangle $ of a particular initial state $\left\vert \psi
(0)\right\rangle $. To utilize the conclusion for the two
sub-systems, the initial state should be chosen as the vacuum state $%
\left\vert \psi (0)\right\rangle =\left\vert \text{\textrm{Vac}}%
\right\rangle $ for both sets of fermion operators $\left\{ \alpha
_{j}\right\} $\ and $\left\{ \beta _{j}\right\} $, i.e.,

\begin{equation*}
\alpha _{j}\left\vert \text{\textrm{Vac}}\right\rangle =\beta _{j}\left\vert 
\text{\textrm{Vac}}\right\rangle =0,
\end{equation*}%
rather than the vacuum state of operator $\left\{ c_{j}\right\} $. The
vacuum state can be constructed as the form%
\begin{equation}
\left\vert \text{\textrm{Vac}}\right\rangle =\prod_{j=1}^{N}\beta _{j}\alpha
_{j}\left\vert 0\right\rangle =\prod_{j=1}^{N}\frac{i}{\sqrt{2}}\left(
1+c_{2j-1}^{\dag }c_{2j}^{\dag }\right) \left\vert 0\right\rangle ,
\end{equation}%
which is essentially the ground state of the Hamiltonian $H$ at the special
point with $\lambda =1$ and $\Delta \ll -1$. Then such a special vacuum
state can be prepared in this way in the experiment.

\begin{figure*}[tbh]
\centering
\includegraphics[width=0.8\textwidth]{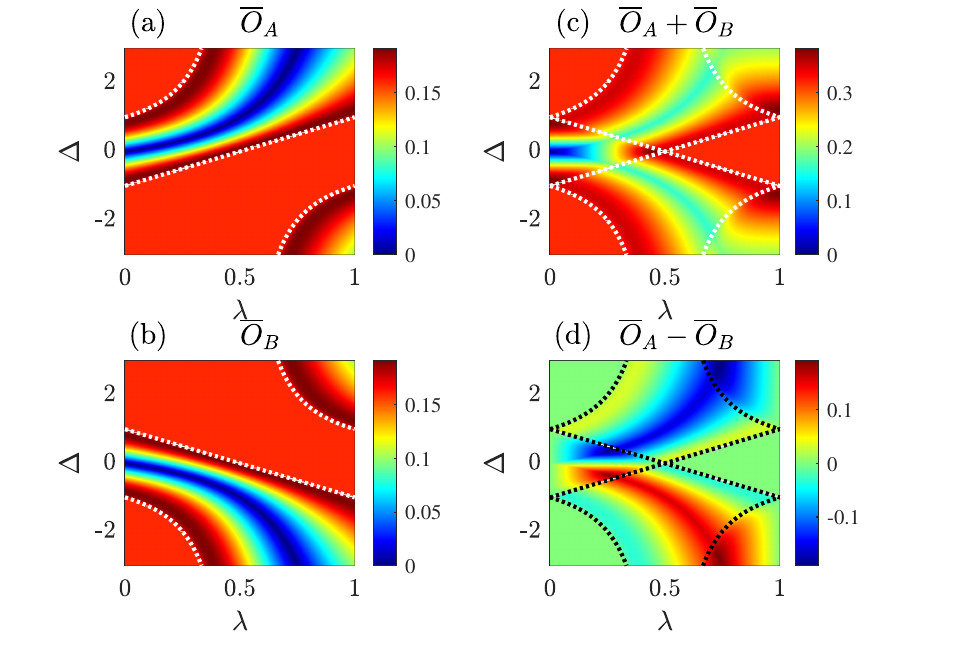}
\caption{Color contour plots of numerical results of dynamic order
parameters (a) $\overline{O}_{A}$ and (b) $\overline{O}_{B}$\ defined in (%
\protect\ref{Oag}) and (\protect\ref{Obg}), respectively. (c) and (d) are
plots of $\overline{O}_{A}+\overline{O}_{B}$ and $\overline{O}_{A}-\overline{%
O}_{B}$, respectively. The system parameters are $N=1000$ and $J=1$. The
white and black dashed lines are a guide to the eye to indicate the phase boundaries
presented in Fig. \protect\ref{fig1}.\ It is clear that the dynamic order
parameters obtained from subsystems can also identify the entire phase
diagram.}
\label{fig3}
\end{figure*}

By replacing the parameters $\left\{ \mu _{A}/J_{A},\mu _{B}/J_{B}\right\} $%
\ by $\left\{ \Delta ,\lambda \right\} $, we obtain the explicit expressions
of $\overline{O}_{A}$\ and $\overline{O}_{A}$\ in the $\lambda \Delta $-plane%
\begin{equation}
\overline{O}_{A}=\frac{1}{2\pi }\left\{ 
\begin{array}{cc}
1, & \text{regions I, II, and III} \\ 
\Omega _{A}, & \text{otherwise}%
\end{array}%
\right.  \label{AveOa}
\end{equation}%
where region I is $\Delta \in (-\frac{1}{2\lambda -1}$, $\infty )$ for $%
0<\lambda <0.5$, region II is $\Delta \in (-\frac{1}{2\lambda -1}$, $%
2\lambda -1)$ for $0.5<\lambda <1$, and region III is $\Delta \in (-\infty $%
, $2\lambda -1)$ for $0<\lambda <0.5$, 
\begin{equation}
\overline{O}_{B}=\frac{1}{2\pi }\left\{ 
\begin{array}{cc}
1, & \text{regions I, II, and III} \\ 
\Omega _{B}, & \text{otherwise}%
\end{array}%
\right.  \label{AveOb}
\end{equation}%
where region I is $\Delta \in (-\infty $, $\frac{1}{2\lambda -1})$ for $%
0<\lambda <0.5$, region II is $\Delta \in (-2\lambda +1$, $\frac{1}{2\lambda
-1})$ for $0.5<\lambda <1$, and region III is $\Delta \in (-2\lambda +1$, $%
\infty )$ for $0<\lambda <0.5$. Here two factors can be expressed explicitly
as%
\begin{widetext}
\begin{equation}
\Omega _{A}=\left\vert \frac{\lambda -\left( 1-\lambda \right) \Delta }{%
\lambda \Delta +\left( 1-\lambda \right) }+\frac{1}{2}\left[ 1-\left( \frac{%
\lambda -\left( 1-\lambda \right) \Delta }{\lambda \Delta +\left( 1-\lambda
\right) }\right) ^{2}\right] \ln \left\vert \frac{2\lambda \Delta +1-\Delta
}{1-2\lambda +\Delta }\right\vert \right\vert ,
\end{equation}%
and%
\begin{equation}
\Omega _{B}=\left\vert -\frac{\lambda +\left( 1-\lambda \right) \Delta }{%
\lambda \Delta -\left( 1-\lambda \right) }+\frac{1}{2}\left[ 1-\left( \frac{%
\lambda +\left( 1-\lambda \right) \Delta }{\lambda \Delta -\left( 1-\lambda
\right) }\right) ^{2}\right] \ln \left\vert \frac{2\lambda \Delta -1-\Delta
}{-1+2\lambda +\Delta }\right\vert \right\vert .
\end{equation}%
\end{widetext}In Fig. \ref{fig3}, we plot the functions, $\overline{O}_{A}$, 
$\overline{O}_{B}$\ and $\overline{O}_{A}\pm \overline{O}_{B}$,
respectively. We can see that quantities $\overline{O}_{A}$ and $\overline{O}%
_{B}$\ reflect partial phase boundary, while both two types of order
parameters $\overline{O}_{A}\pm \overline{O}_{B}$\ can identify the entire
phase diagram.\ Methodologically, it also demonstrates the benefits of the
real-space decomposition.

\begin{figure*}[tbh]
\centering
\includegraphics[width=0.8\textwidth]{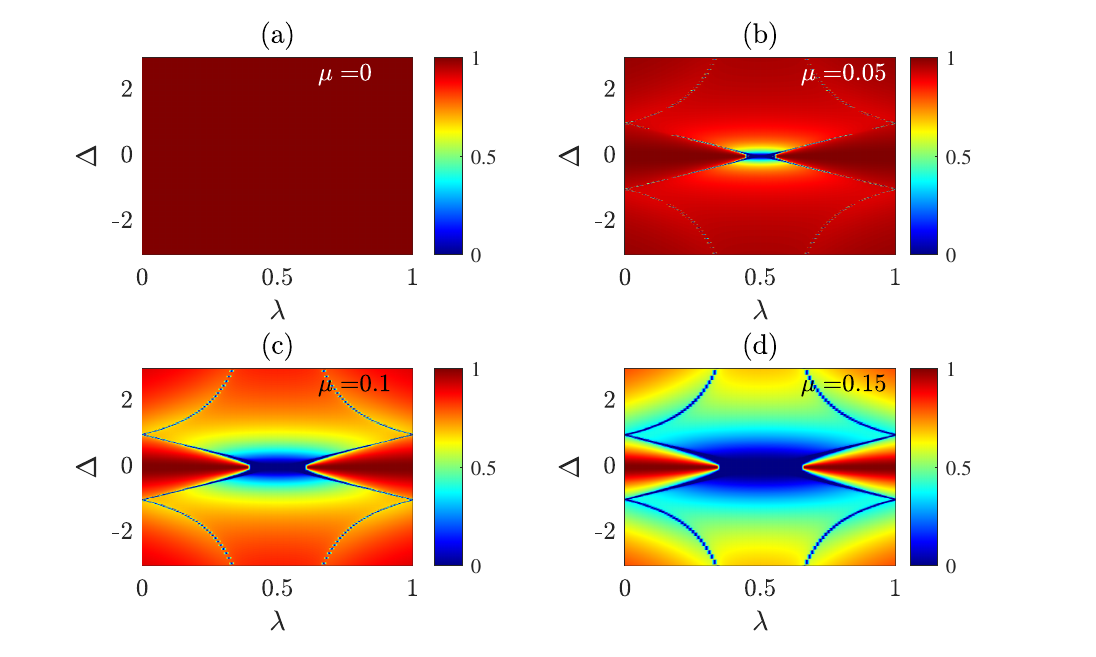}
\caption{The plots of the fidelity of $f\left( \protect\lambda ,\Delta
\right) $ defined in (\protect\ref{fidelity}) for several representative
values of $\protect\mu $, (a) $\protect\mu =0$, (b) $\protect\mu =0.5$, (c) $%
\protect\mu =1.0$, and (d) $\protect\mu =1.5$. The results are obtained by
exact diagonalization for the Hamiltonians with $N=100$. It shows the ground
state remains unchanged within many gapful regions, indicating that the
real-space decomposition for the ground state still holds true approximately
in the presence of finite $\protect\mu $.}
\label{fig4}
\end{figure*}

\section{Robustness against nonzero chemical potential}

\label{Robustness against nonzero}

The real-space decomposition is exact when the chemical potential is zero.
In this section, we investigate the influence\ of nonzero $\mu $ on our
above conclusions. The decomposition we proposed is no longer valid, in the
presence of nonzero $\mu $. However, we will show that our conclusion for
the ground state holds approximately in most of regions for small values of $%
\mu $. In principle, when small $\mu $\ switches on, one can treat $H_{AB}$\
as a perturbation term. The perturbation theory tells us the effect of the
perturbation term depends strongly on the eigenstates of $H_{A}+H_{B}$. In
the limit case, the energy gaps for the groundstate of both $H_{A}+H_{B}$
are sufficiently large comparing to the value of $\mu $, $H_{AB}$\ should
have no effect on the groundstates $\left\vert \text{\textrm{G}}%
_{A}\right\rangle \left\vert \text{\textrm{G}}_{B}\right\rangle $\ of $%
H_{A}+H_{B}$. On the other hand, at the phase boundary, the gapless point,
the ground states $\left\vert \text{\textrm{G}}_{A}\right\rangle \left\vert 
\text{\textrm{G}}_{B}\right\rangle $ become (or quasi-) degenerate. The term 
$H_{AB}$\ may hybridize them and the low-lying excited states, resulting in
entangled state, which is deviated from the product state $\left\vert \text{%
\textrm{G}}_{A}\right\rangle \left\vert \text{\textrm{G}}_{B}\right\rangle $.

Based on the Fourier transformations for two sub-lattices in Eq. (\ref%
{Fourier}), the Hamiltonian with periodic boundary condition can be block
diagonalized by this transformation due to its translational symmetry, i.e.,%
\begin{equation}
H=\sum_{k\in \left[ 0,\pi \right] }H_{k}=H_{0}+H_{\pi }+\sum_{k\in (0,\pi
)}\psi _{k}^{\dagger }h_{k}\psi _{k},
\end{equation}%
satisfying $\left[ H_{k},H_{k^{\prime }}\right] =0$, where the operator
vector $\psi _{k}^{\dagger }=\left( A_{k}^{\dagger },B_{k}^{\dagger
},A_{-k},B_{-k}\right) $, and the core matrix is expressed explicitly as

\begin{equation}
h_{k}=\left( 
\begin{array}{cccc}
\mu & z & 0 & w \\ 
z^{\ast } & \mu & -w^{\ast } & 0 \\ 
0 & -w & -\mu & -z \\ 
w^{\ast } & 0 & -z^{\ast } & -\mu%
\end{array}%
\right) ,
\end{equation}%
where $z=\lambda e^{ik}$ $+\left( 1-\lambda \right) $ and $w=\Delta \left[
\lambda \text{ }-\left( 1-\lambda \right) e^{ik}\right] $. Here, $H_{0}$ and 
$H_{\pi }$ have the form%
\begin{eqnarray}
H_{0} &=&2JA_{0}^{\dagger }B_{0}+2JB_{0}^{\dagger }A_{0}+2\mu \left(
A_{0}^{\dag }A_{0}-B_{0}B_{0}^{\dag }\right)  \notag \\
&&+\left( \Delta _{a}-\Delta _{b}\right) \left( A_{0}^{\dagger
}B_{0}^{\dagger }+A_{0}B_{0}\right) , \\
H_{\pi } &=&2\mu \left( A_{\pi }^{\dag }A_{\pi }-B_{\pi }B_{\pi }^{\dag
}\right)  \notag \\
&&-\left( \Delta _{a}+\Delta _{b}\right) \left( A_{\pi }^{\dagger }B_{\pi
}^{\dagger }+B_{\pi }A_{\pi }\right) ,
\end{eqnarray}%
and $H_{\pi }$\ vanishes when odd $N$ is taken. To demonstrate the above
analysis, one can rewrite the matrix $h_{k}$\ in the form 
\begin{equation}
h_{k}=h_{k}^{0}+\mu \Gamma ^{z}
\end{equation}%
with

\begin{equation}
h_{k}^{0}=\left( 
\begin{array}{cccc}
0 & z & 0 & w \\ 
z^{\ast } & 0 & -w^{\ast } & 0 \\ 
0 & -w & 0 & -z \\ 
w^{\ast } & 0 & -z^{\ast } & 0%
\end{array}%
\right) ,
\end{equation}%
and a $k$-independent matrix%
\begin{equation}
\Gamma ^{z}=\left( 
\begin{array}{cccc}
1 & 0 & 0 & 0 \\ 
0 & 1 & 0 & 0 \\ 
0 & 0 & -1 & 0 \\ 
0 & 0 & 0 & -1%
\end{array}%
\right) .
\end{equation}%
The eigen values $\varepsilon _{\sigma \rho }^{k}$ and vectors $\phi
_{\sigma \rho }^{k}$ of $h_{k}^{0}$\ can be obtained as%
\begin{equation}
\left( 
\begin{array}{c}
\left( \phi _{++}^{k}\right) ^{T} \\ 
\left( \phi _{+-}^{k}\right) ^{T} \\ 
\left( \phi _{-+}^{k}\right) ^{T} \\ 
\left( \phi _{--}^{k}\right) ^{T}%
\end{array}%
\right) =\frac{1}{2}\left( 
\begin{array}{cccc}
\eta _{k} & 1 & -\eta _{k} & 1 \\ 
\eta _{k} & -1 & \eta _{k} & 1 \\ 
\eta _{k} & -1 & -\eta _{k} & -1 \\ 
\eta _{k} & 1 & \eta _{k} & -1%
\end{array}%
\right) ,
\end{equation}%
and%
\begin{equation}
\varepsilon _{\sigma \rho }^{k}=\sigma \left\vert w+\rho z\right\vert 
\end{equation}%
satisfying $h_{k}^{0}\phi _{\sigma \rho }^{k}=\varepsilon _{\sigma \rho
}^{k}\phi _{\sigma \rho }^{k}$, where $\eta _{k}=\frac{w+z}{\left\vert
w+z\right\vert }$, with the indices $\sigma
,\rho =\pm $.\ It is easy to check that the $h_{k}^{\mu }$\ acts as a flip
operator%
\begin{equation}
\Gamma ^{z}\phi _{\sigma \rho }^{k}=\phi _{\overline{\sigma }\overline{\rho }%
}^{k},
\end{equation}%
with the labels $\left( \overline{\sigma },\overline{\rho }\right) =\left(
-\sigma ,-\rho \right) $. It directly results in the transition matrix
element between energy bands 
\begin{equation}
\left( \phi _{\sigma \rho }^{k}\right) ^{\dag }\Gamma ^{z}\phi _{\sigma
^{\prime }\rho ^{\prime }}^{k}=\delta _{\sigma ^{\prime }\overline{\sigma }%
}\delta _{\rho ^{\prime }\overline{\rho }}.
\end{equation}%
This indicates that when the energy gap is sufficiently large compared to the
values of the chemical potential,%
\begin{equation}
\left\vert \mu \right\vert \ll \left\vert \varepsilon _{\sigma \rho
}^{k}-\varepsilon _{\overline{\sigma }\overline{\rho }}^{k}\right\vert ,
\end{equation}%
the term $\mu \Gamma ^{z}$ should not affect the eigen vectors of $h_{k}^{0}$%
.

We introduce the quantity%
\begin{equation}
f\left( \lambda ,\Delta \right) =\left\vert \langle \text{\textrm{G}}(\mu
)\left\vert \text{\textrm{G}}_{A}\right\rangle \left\vert \text{\textrm{G}}%
_{B}\right\rangle \right\vert ^{2}=\left\vert \langle \text{\textrm{G}}(\mu
)\left\vert \text{\textrm{G}}(0)\right\rangle \right\vert ^{2},
\label{fidelity}
\end{equation}%
to quantitatively measure the effect of $\mu $\ on the groundstate $%
\left\vert \text{\textrm{G}}\right\rangle $\ of $H$. Our above analysis
predicts that the gapless lines of both $H_{A}$ and $H_{B}$\ result in the
valley of $f\left( \lambda ,\Delta \right) $. Numerical simulations for $%
f\left( \lambda ,\Delta \right) $\ in finite size system can be performed by
exact diagonalization. It essentially relates to two ground states of $H$ with
zero and nonzero $\mu $, respectively, while $\left\vert \text{\textrm{G}}%
(0)\right\rangle =\left\vert \text{\textrm{G}}_{A}\right\rangle \left\vert 
\text{\textrm{G}}_{B}\right\rangle $ is the ground state for zero $\mu $.
For given parameters, $f\left( \lambda ,\Delta \right) $ can be obtained by
diagonalization of matrix $h_{k}$ and the eigenstates of $H_{0}+H_{\pi }$.

Fig. \ref{fig4} shows the fidelity of $f\left( \lambda ,\Delta \right) $ for
a given finite system as a function of $\left( \lambda ,\Delta \right) $
with various values of $\mu $. In the case of zero $\mu $, the fidelity is
unitary everywhere as expected. In plot for small $\mu $, there are
evidently sudden drops of the value of fidelity at the phase boundaries and
the drops become sharper and sharper as $\mu $\ increases. These behaviors
can be ascribed to a dramatic change of the ground state of the system
around the phase boundaries. One can also find out that the ground state
remains unchanged within many regions. It indicates the real-space
decomposition for the ground state still holds true approximately in many
regions in the presence of finite $\mu $.

\section{Summary}

\label{sec_summary}

In summary, we have extended the Bogoliubov transformation for spinless
fermions in $k$-space to the one in real space, which has been shown to be a
tool for the spatial decomposition of a class of Kitaev chains. It provides
a way to get insight into a complicated system from that of two decoupled
sub-systems. A systematic investigation of a SSH Kitaev model is performed,
including the ground state property and nonequilibrium behavior of quenching
dynamics. In addition, we also studied the approximate decomposition in
the case of nonzero chemical potential. Analytical analysis and numerical
simulation show that the real-space decomposability is maintained in
most regions of the phase diagram when the chemical potential is small. Our
findings not only contribute to the methodology for solving the many-body
problem, but also reveal the underlying mechanism of the features of the
SSH Kitaev chain.

\section*{Acknowledgment}

We acknowledge the support of NSFC (Grants No. 12374461).

\end{document}